\input harvmac

\Title{\vbox{\baselineskip12pt
\hbox{BCCUNY-HEP/02-03} 
\hbox{hep-th/0209049}}}
{\vbox{\centerline{Remarks on Fundamental String Cosmology}}}

\baselineskip=12pt \centerline {Ramzi R. Khuri\footnote{$^*$}
{e-mail: khuri@gursey.baruch.cuny.edu.}}
\medskip
\centerline{\sl Department of Natural Sciences, Baruch College, CUNY} 
\centerline{\sl 17 Lexington Avenue, New York, NY 10010}
\medskip
\centerline{\sl Graduate School and University Center, CUNY}
\centerline{\sl 365 5th Avenue, New York, NY 10036}

\bigskip
\centerline{\bf Abstract}
\medskip
\baselineskip = 20pt

In recent work, it was shown that velocity-dependent forces
between moving strings or branes lead to an accelerating expanding
universe without assuming the existence of a cosmological constant. Here 
we show that the repulsive velocity-dependent force arises in more general
contexts and can lead to cosmic structure formation.

\Date{September 2002}

\def\r{\rho}

\def\({\left (}
\def\){\right )}
\def\[{\left [}
\def\]{\right ]}

\lref\fund{R. R. Khuri, Phys. Lett. {\bf B520} (2001) 353, hep-th/0109041.}

\lref\shell{R. R. Khuri and A. Pokotilov, Nucl. Phys. {\bf B633} (2002)295, hep-th/0201194;
Phys. Lett. {\bf B535} 1 (2002) hep-th/0202158.}

\lref\ferr{R. C. Ferrell and D. M. Eardley, Phys. Rev. Lett. {\bf 59} (1987) 1617.}

\lref\prep{M. J. Duff, R. R. Khuri and J. X. Lu, Phys. Rep.
{\bf B259} (1995) 213, hep-th/9412184.}

\lref\bps{M. K. Prasad and C. M. Sommerfield, Phys. Rev. Lett. {\bf 35}
(1975) 760; E. B. Bogomol'nyi, Sov. J. Nucl. Phys. {\bf 24} (1976) 449.}

In \refs{\fund,\shell}, it was shown that the velocity-dependent 
forces between 
parallel fundamental strings lead to an expanding universe in one less
dimension. The same result was noted to hold for parallel 
identical $p$-branes in general as well as for dual branes, such as
the case of fundamental strings propagating in a solitonic fivebrane 
background. In the particular case of a four-dimensional universe,
it was shown by both a mean-field approximation \fund\ and a spherical
``shell" model \shell\ that the early universe for these models undergoes
an inflationary exponential growth, equivalent to a positive cosmological
constant. Power-law accelerating expansion was found for other 
dimensions.

The expanding spherical model of \shell\ is easily adaptable to the
case of a spatially flat ($k=0$) universe. Following the calculation of \shell,
one again finds an exponential growth in the early universe. This is certainly no surprise,
since the mean-field approximation of \fund, whose findings agree with
those of \shell, makes no reference to the curvature of the universe.
We shall therefore assume henceforth that the mean-field approximation 
represents a reasonable first-approximation to the general expansion 
of the universe in this kind of model.

The mean-field approximation essentially assumes that the many-body
dynamics of the moving strings can be averaged by the one-body
problem of a single string moving in the background of a much larger
string. Let $x^i$ be the coordinate of the moving test string in the
transverse space, where the
source string is at the origin, $r^2 = x^i x_i$ and
$\dot x^2 = \dot x^i \dot x_i$. Let $h=1+ k/r^n$, where $k$ is the
Noether charge of the source string, and $n=D-4$, where $D$ is the 
number of spacetime dimensions. We obtain the Lagrangian \fund
\eqn\lag{ {\cal L} = -m h^{-1} \left ( \sqrt{1- h \dot x^2} -1 \right),}
and Hamiltonian
\eqn\ham{H = {m\over h} \left( {1\over \sqrt{1-h \dot x^2}}-1\right)= E,}
where $m$ is the mass of the test string and $E$ is its constant energy.
For parallel fundamental strings in $D$ spacetime dimensions, \lag\ and \ham\
lead to an accelerating, expanding universe. 

The spherically-symmetric  multi-string model in \shell\ was constructed
to test the validity of the mean-field results, with agreement being
found between the two methods.

The exponential accelerating universe found in \refs{\fund,\shell}\
is not peculiar to string theory. It can easily be checked that the
low-velocity Lagrangian for, say, moving extremal Reissner-Nordstr\" om
black holes \ferr\ also leads to an exponentially expanding universe
in the mean-field approximation. In this case, the Lagrangian is a
low-velocity two-body interaction
\eqn\ferlag{ L = -M + {1\over 2} M {\vec V}^2 + {1\over 2} \mu {\vec v}^2
\left( \left[ 1 + M/r \right]^3 -2\mu M^2/r^3 \right),}
where $M= m_1 +m_2$, $\mu = m_1 m_2 /M$, $\vec V$ is the CM velocity,
$\vec v$ is the relative velocity and $r= | \vec r |$ is the relative separation
between the two black holes.
In fact, it was shown in \ferr\ that three-body and four-body interactions also exist, 
but these lead to the same result in the mean-field limit for the general 
behaviour of the expansion.

One can go even further and show that the mean-field approximation
of \fund\ leads to an accelerating expanding universe whenever
the velocity-dependent Lagrangian has the form
\eqn\lagone{L = {m v^2 \over 2} \left( 1 + f(r)v^2 \right),}
as in the case of maximally supersymmetric strings or branes up to
$O(v^4)$ in \refs{\fund,\shell}, or of the form
\eqn\lagtwo{L = {m v^2 \over 2} \left( 1 + g(r) \right),}
as in \ferr, whenever $f(r)$ and $g(r)$ are monotonically decreasing functions of
the separation $r$. Assuming only radial motion, this can be easily seen from
the Euler-Lagrange equations by noting that the radial acceleration $\ddot r$ is always
positive, given by
\eqn\accone{\ddot r = -{3 \dot r^4 f' \over 2(1+6f\dot r^2)}}
for \lagone\ and by
\eqn\acctwo{\ddot r = -{\dot r^2 g' \over 2(1+g)}}
for \lagtwo, where $'$ indicates differentiation with respect to $r$, so that
$f'$ and $g'$ are always negative.

The fact that this kind of velocity-dependent force is always repulsive easily
explains an accelerating growth in the size of the model universe, simply since
it implies an accelerating separation between the various constituents. This kind of
interaction, on the other hand, would
seem to make it difficult, if not impossible, for any cosmic structure, or clumping
of any sort, to arise. Interestingly enough, this may not be the case at all.

This can be seen by revisiting, say, the fundamental string model of \lag.
It was shown in \fund\ that, if an initial outward motion is assumed, then
by using a mean-field approximation, one can replace the total repulsive force by
a single radial force due to a large source string at the origin. From \ham, it follows
that
\eqn\rdotgen{ \dot r^2 = {\rho \left( h\rho +2 \right) \over \left( h\rho + 1 \right)^2},}
where $\rho = E/m$. Here $k$ essentially represents the total Noether charge of these
extremal fundamental strings and therefore the total  mass of the universe in this model.
For $n=2$, in the limit of small $\dot r$, \rdotgen\ reduces to
\eqn\rsmallv{\dot r^2 \simeq {\r^2\over k},}
which for initial positive $\dot r$ leads to the exponential growth
\eqn\expplus{r = r_0 \exp{(t/\sqrt{k})}.}

Suppose now that, within the exponential expansion in the early universe of the
fundamental string (or brane) model of \lag, a test string had an initial outward
velocity, but at some point acquired a velocity {\it towards} a local configuration of
strings, whose total mass is much larger than that of the test string. 
Again, in the mean-field approximation, we assume that the repulsive force
due to this local mass configuration may be taken to be radial, governed by the same
equation \rdotgen, but now with effective Noether charge $k' < k$. Since the test
string is oriented towards the local configuration, the solution of \rsmallv\
for the separation $r'$ between the test string and the local configuration
takes the form 
\eqn\expminus{r' = r'_0 \exp{(-t/\sqrt{k'})}.}
Thus the velocity-dependent force, while still repulsive, allows for the test string
to head towards the local configuration, essentially acting as a critical damping.
This allows for structure formation within the general expansion.
The same result occurs for Lagrangians \lagone\ and \lagtwo.

To summarize, the strings may be moving outwardly in general, at an accelerating pace,
but the nonradial components of the velocities necessarily lead to 
some strings moving towards each other. The velocity dependent forces between the
string is at all times repulsive, but it is such that the strings may yet join together,
or produce stable clusters and hence cosmic structure. This scenario occurs
in the very earliest stages of the expansion, after which the whole model probably
loses validity, but once the anisotropies (or density variations) are formed, they can presumbly
be stabilized, once gravitational forces begin to outweigh the repulsive quantum
interactions. This might in turn occur as a result of the departure of the strings 
from extremality due to the acquisition of energy. 

The above arguments are based on a mean-field approximation, which is assumed to be
fairly robust. The specific toy model considered here is much too simplistic to represent a serious
cosmological model. In particular, in the absence of specific assumptions on the anisotropy
arising from the departures from a purely radial expansion, the time scale
for structure formation implied by \expminus\ is much too small \foot{This was pointed out
to me by Dan Chung.}. Nevertheless, the possibility that clumping could take place in the context
of the repulsive velocity-dependent force suggests that a more realistic model of this type
may give rise to structure formation. It would therefore be of interest to try to understand 
the general statistical multi-body problem for these interacting strings or extremal black holes.
Work on this problem is in progress.

{\bf Acknowledgements:} I would like to thank the CERN Theory Division for their hospitality.
I would like to thank Dan Chung, Elias Kiritsis, Costas Kounnas, Fernando Quevedo and Gabrielle 
Veneziano for helpful discussions.

\listrefs

\end